\documentclass[journal=aelccp,manuscript=letter,layout=traditional]{achemso} 

\usepackage[version=3]{mhchem} 

\usepackage{hyperref}
\usepackage{graphicx}
\usepackage{subcaption}
\usepackage{xr}

\author{Thijs J. A. M. Smolders}
\affiliation{Department of Physics, University of Bath, Claverton Down, Bath, BA2 7AY, United Kingdom}
\author{Alison B. Walker}
\affiliation{Department of Physics, University of Bath, Claverton Down, Bath, BA2 7AY, United Kingdom}
\author{Matthew J. Wolf}
\affiliation{Department of Physics, University of Bath, Claverton Down, Bath, BA2 7AY, United Kingdom}
\email{m.j.wolf@bath.ac.uk}

\title{3D-to-2D Transition of Anion Mobility in \ce{CsPbBr3} under Pressure}

\begin{document}
\begin{abstract}
We study the effects of hydrostatic pressure in the range 0.0--2.0 GPa on anion mobility in the orthorhombic $Pnma$ phase of \ce{CsPbBr3}. Using density functional theory and the climbing nudged elastic band method, we calculate the transition states and activation energies for anions to migrate both within and between neighbouring \ce{PbBr3} octahedra. The results of those calculations are used as input to a kinetic model for anion migration, which we solve in the steady state to determine the anion mobility tensor as a function of applied pressure. We find that the response of the mobility tensor to increasing pressure is highly anisotropic, being strongly enhanced in the $(010)$ lattice plane and strongly reduced in the direction normal to it at elevated pressure. These results demonstrate the potentially significant influence of pressure and strain on the magnitude and direction of anion migration in lead--halide perovskites.
\end{abstract}
\begin{figure}[h]
\centering
\begin{subfigure}{0.9\textwidth}
\centering
    \includegraphics[width=\linewidth]{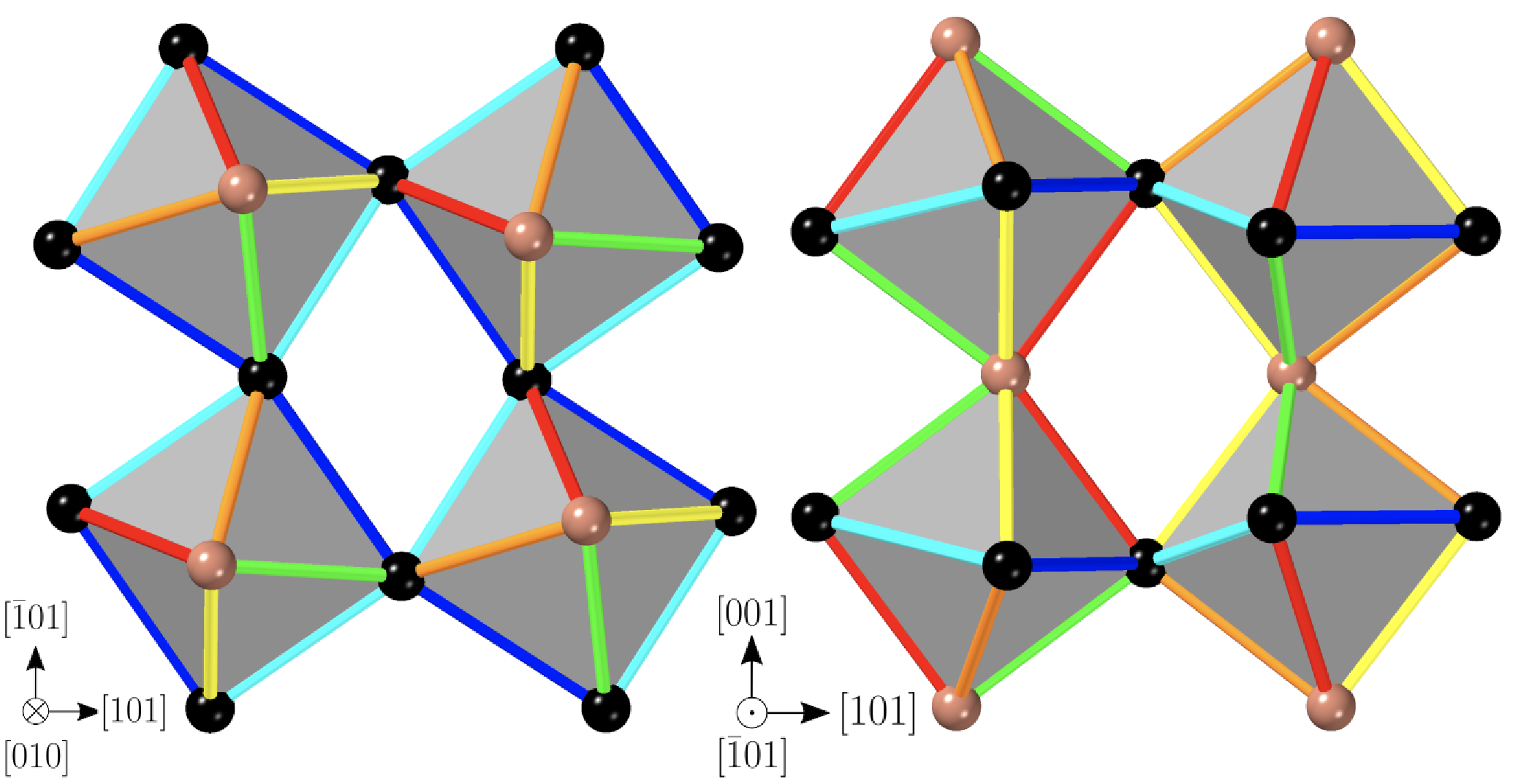}
    \caption*{}
    \label{fig:paths_NN_2x2_v1}
\end{subfigure}%

\begin{subfigure}{0.45\textwidth}
\centering
    \includegraphics[width=\linewidth]{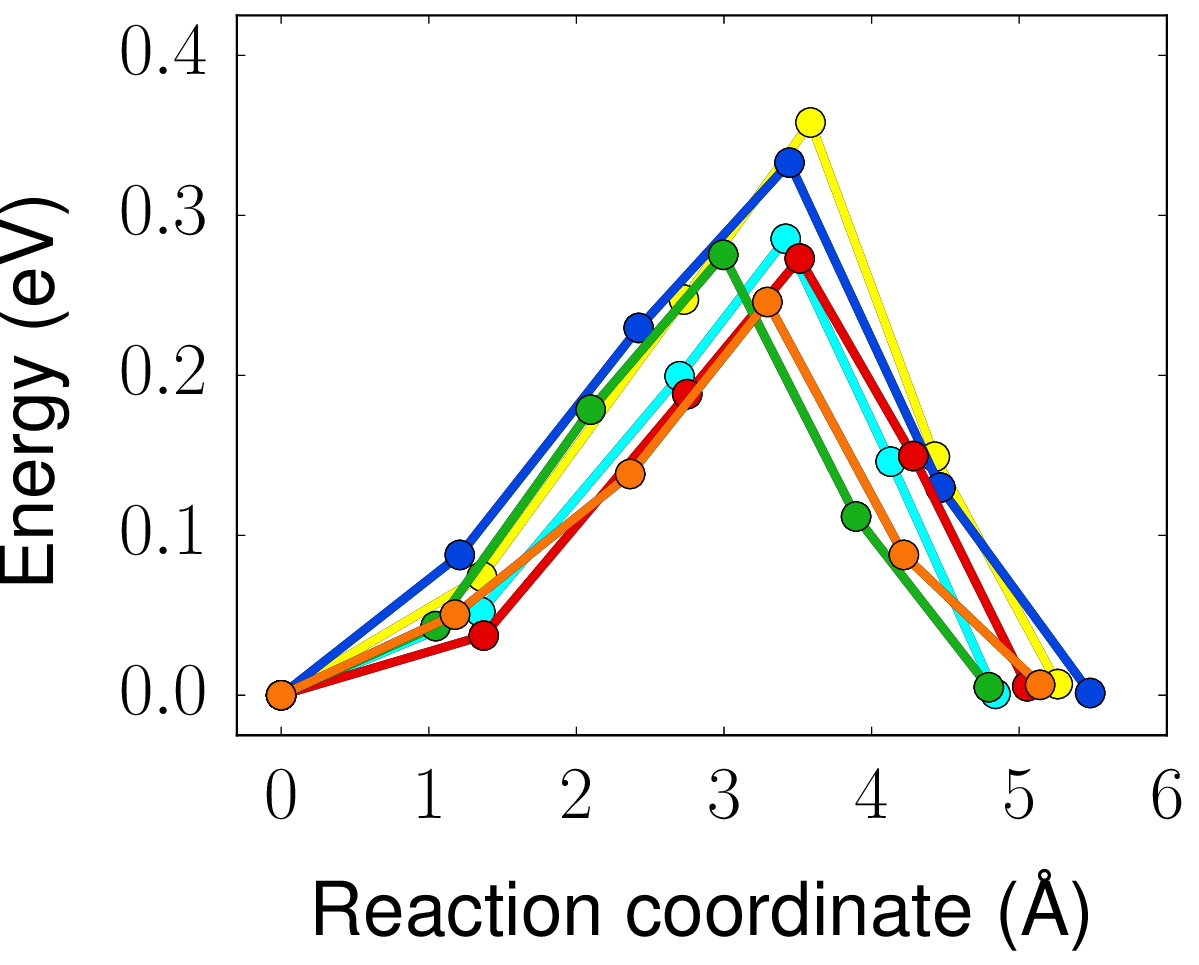}
    \caption*{}
    \label{fig:mig_bars_0_GPa}
\end{subfigure}%
\hfill
\begin{subfigure}{0.45\textwidth}
\centering
    \includegraphics[width=\linewidth]{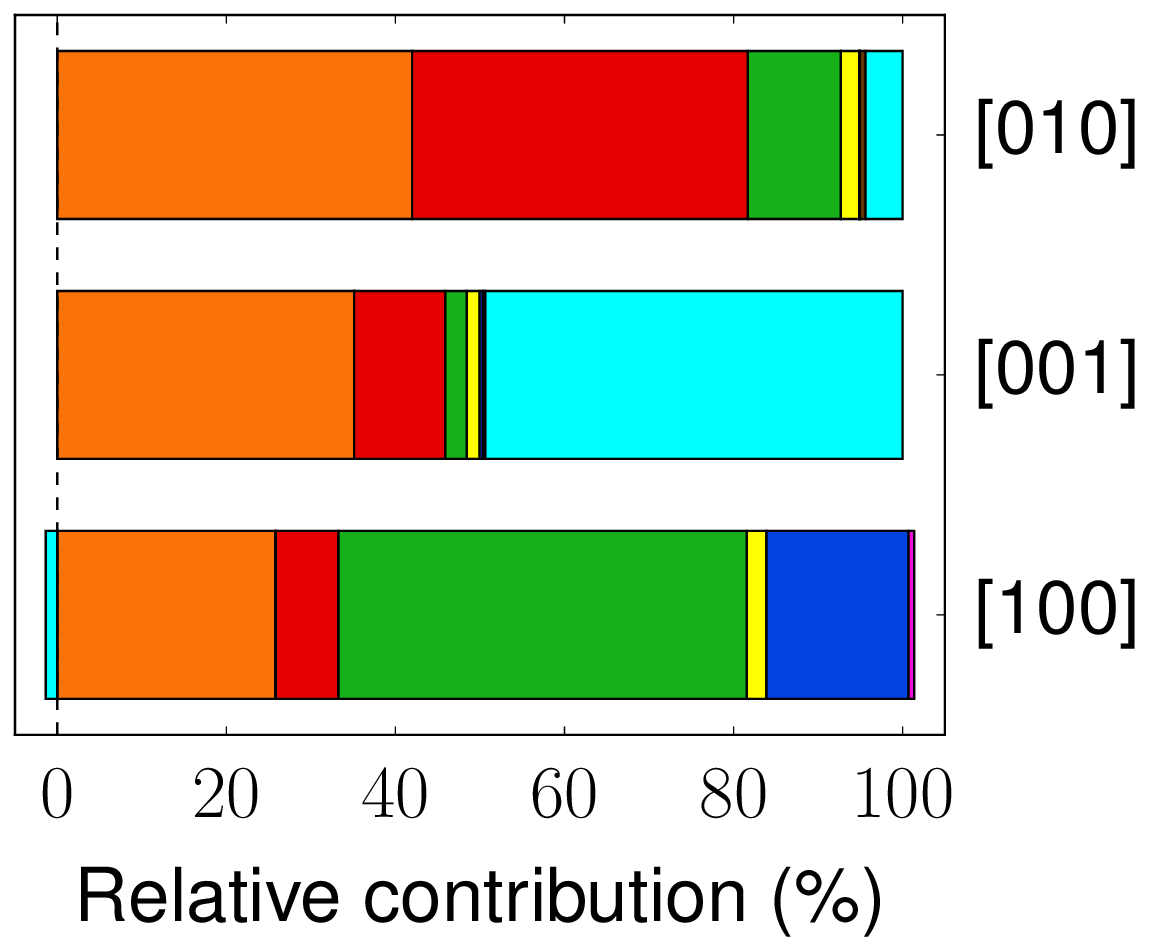}
    \caption*{}
    \label{fig:decomp_0_GPa}
\end{subfigure}
\caption{Top row: Structure of the \textit{Pnma} phase of \ce{CsPbBr3} at 0.0 GPa. Only the bromide ions are shown, with the symmetrically inequivalent apical and equatorial bromide sites represented by pink and black spheres, respectively. The symmetrically inequivalent \emph{pairs} of 1-NN bromide ions are connected along the edges of the \ce{PbBr3} octahedra by colour-coded sticks.  On the left, the structure is viewed along the \textbf{b} (long) axis, thus highlighting the connectivity in the equatorial plane, while on the right, the structure is viewed along the [$\bar{1}$01] direction, highlighting the connectivity along the \textbf{b} axis, which points up the page. Bottom left; the energy barriers for positively charged bromide vacancy migration along the symmetrically inequivalent paths, at 0.0 GPa. Bottom right; the relative contributions of transitions between each symmetrically inequivalent pair to the overall mobility, at 0.0 GPa.}
\label{fig:disp_vecs_0_GPa}
\end{figure}

Over the past decade, lead--halide perovskites (LHPs) have emerged as promising materials for optoelectronic applications, most notably as active layers in photovoltaic cells \cite{Snaith2018a, Jena2019} and LEDs \cite{Service2019,Lu2019,Liu2020b}. One of the advantageous qualities of LHPs is that their properties may be readily tuned, which has most notably been achieved by altering the composition \cite{Protesescu2015,Jeon2015,Saliba2016a,Li2019e}. More recently, strain engineering has been suggested as an additional means of modifying the properties of LHPs \cite{Zhu2019a,Kim2020c}. Residual levels of strain, resulting from lattice mismatch and thermal expansion mismatch, are known to strongly affect LHP properties \cite{Jones2019}. In particular, enabled by the softness of the materials \cite{Sun2015c,Letoublon2016,Lee2018c,Ferreira2018}, pressure offers a particularly powerful way of inducing strain \cite{Jaffe2017b,Li2019d,Liu2019b,Jiao2020}.
\\
The effects of pressure on the electronic properties of LHPs have been the subject of a number of experimental studies \cite{Jaffe2017a,Lu2017,Li2020}, and theoretical calculations have provided explanations for the observed behaviour in terms of the different modes of distortion of the lattice under pressure, and their effects on inter-ionic bond lengths and bond angles \cite{Xiao2017c,Huang2019,Wang2019i,Wang2019e,Ghosh2019a,Li2020b}.
\\
In addition to electronic transport, LHPs are also well known to exhibit significant ionic conductivity \cite{Pockett2017,Tress2017b,Walsh2018,Lee2019a} even at moderate temperatures, and moreover, ion transport has been implicated in a number of critical aspects of device performance, such as current--voltage hysteresis \cite{Richardson2016,Habisreutinger2018} and the long-term stability \cite{Yuan2016,Walsh2018,Zhang2019_rev,Lee2019a,Zhao2021} of LHP-based devices. Furthermore, a number of recent studies have indicated that ionic conductivity in LHPs is also significantly affected by various manifestations of strain, both residual \cite{Zhao2017b,Xue2020} and pressure-induced \cite{Ou2018a,Ou2019,Muscarella2020}. However, unlike the electrical conductivity, a detailed atomistic understanding of the influence of pressure on ion motion is currently lacking.
\\
In this work, we carry out a quantitative analysis of the effects of pressure on ionic conductivity in the all-inorganic LHP CsPbBr$_{3}$. We compute migration barriers using density functional theory (DFT) and the climbing nudged elastic band method \cite{Henkelman2000,Sheppard2008}, and use the resulting activation energies as parameters in a kinetic scheme \cite{Banyai1979,Allnatt2003} that we solve in the steady state to obtain a macroscopic anion mobility tensor as a function of applied pressure. \ce{CsPbBr3} is a promising material for a number of applications, such as LEDs and photodetectors \cite{Kovalenko2017,Shen2020,Yu2020}, although its band gap is generally considered to be too wide for applications in single junction solar cells. It adopts the same orthorhombic $Pnma$ structure at low temperatures as other techonologically relevant LHPs, such as \ce{MAPbI3}, \ce{MAPbBr3} and \ce{FAPbBr3} \cite{Lee2016a,Bernasconi2017,Schueller2018}; however, unlike those materials, it does not undergo a phase transition until well above room temperature \cite{Stoumpos2013a,Malyshkin2020}. It is important to note that density functional theory (DFT) total energy calculations effectively correspond to a temperature of 0 K, at which the commonly considered higher symmetry phases, such as the tetragonal \textit{P4/mbm} and cubic \textit{Pm}$\bar{3}$\textit{m} phases, are unstable \cite{Whalley2016,Yang2017c,Marronnier2017}; additionally, several studies have suggested that these higher symmetry phases, which are typically observed at higher temperatures in halide perovskites, are in fact dynamical averages of the orthorhombic \textit{Pnma} phase \cite{Bechtel2018a,Klarbring2019,Bechtel2019}, making \ce{CsPbBr3} a good model system for the general class of LHPs.  
\\
Anion vacancies are believed to be the most mobile species in LHPs in general \cite{Eames2015,Haruyama2015,Azpiroz2015,Meloni2016,Mosconi2016,Luo2017,Chen2019c,Zhang2020d}, and the positive charge state of the bromide vacancy in \ce{CsPbBr3} has been shown to be the most stable one for Fermi levels within the majority of the band gap \cite{Shi2014,Sebastian2015,Kang2017a,Kang2020}; therefore, we restrict our study to the migration of positively charged bromide vacancies.
\\
Perovskite structures of stoichiometry \ce{ABX3} are formed of corner sharing octahedra of X anions, with B cations at their centres and A cations in the voids between them. The perovskite structure of highest symmetry, the so-called ``aristotype'', is a cubic structure of space group \textit{Pm}$\bar{3}$\textit{m}, with a unit cell that contains a single \ce{ABX3} formula unit. Lower symmetry perovskite structures are thought of primarily as resulting from taking the aristotype and tilting neighbouring octahedra with respect to one another. In the orthorhombic \textit{Pnma} structure, the tilting pattern corresponds to in--phase tilts about the long (\textbf{b}) lattice vector, and out--of--phase tilts along the other two lattice vectors ($a^{-}b^{+}a^{-}$ in the notation of Glazer \cite{Glazer1972}), resulting in a unit cell which contains four \ce{ABX3} formula units.
\\
The lower symmetry of the orthorhombic structure with respect to the cubic one has important implications for the number of elementary anion migration paths connecting pairs of lattice sites that must be considered \cite{Gao2020}. Firstly, while all anion sites are symmetrically equivalent in the cubic structure, in the orthorhombic structure there are two symmetrically inequivalent sites, corresponding to the 4c and 8d Wyckoff positions, which we describe as being either apical or equatorial respectively, and which are shown in Figure \ref{fig:disp_vecs_0_GPa} as either pink or black spheres, respectively. Thus, each migration event proceeds from one of two types of site. Secondly, while each of the eight first-nearest neighbours (1-NNs) of a given halide site, which correspond to the final site of a migration event, are symmetrically equivalent (with respect to the point group of the site) in the cubic structure, this is not the case in the orthorhombic structure, as we will now discuss in more detail.
\\
We first consider the point group symmetry at the apical site. All of its 1-NN bromide sites are of the equatorial type, and belong to the two octahedra that are corner-sharing through the apical bromide. By virtue of the mirror symmetry at the apical site, these octahedra are symmetrically equivalent, and therefore, the eight 1-NNs can be grouped into four symmetrically inequivalent pairs. Next, we look at the local symmetry at the equatorial position. Of its eight 1-NNs, four are apical and the relation between them has already been discussed above. This leaves four remaining 1-NNs, all of which are equatorial, and a screw symmetry axis reduces the remaining four neighbours to two symmetrically inequivalent pairs. A visual representation of the resulting six symmetrically inequivalent pairs of initial and final bromide sites involved in elementary migration events is shown in Figure \ref{fig:disp_vecs_0_GPa}.

To find the migration paths and activation energies for each of the elementary migration events (defined by the symmetrically inequivalent pairs of initial and final bromide lattice sites), climbing nudged elastic band (CNEB) calculations, based on density functional theory, were carried out. The results of the calculations are shown in Figure \ref{fig:disp_vecs_0_GPa}. Relative to the aristotype, the reduced symmetry of the \textit{Pnma} stucture results in a significant spread in barrier heights for the different transitions. The magnitudes, in the range of 0.23--0.34 eV, are in line with values reported in experimental studies \cite{Mizusaki1983,Narayan1987} and theoretical calculations of ion migration in the higher symmetry tetragonal and cubic phases \cite{Meloni2016,Muscarella2020}.

\begin{figure}[h]
\centering
\begin{subfigure}{0.9\textwidth}
\centering
    \includegraphics[width=\linewidth]{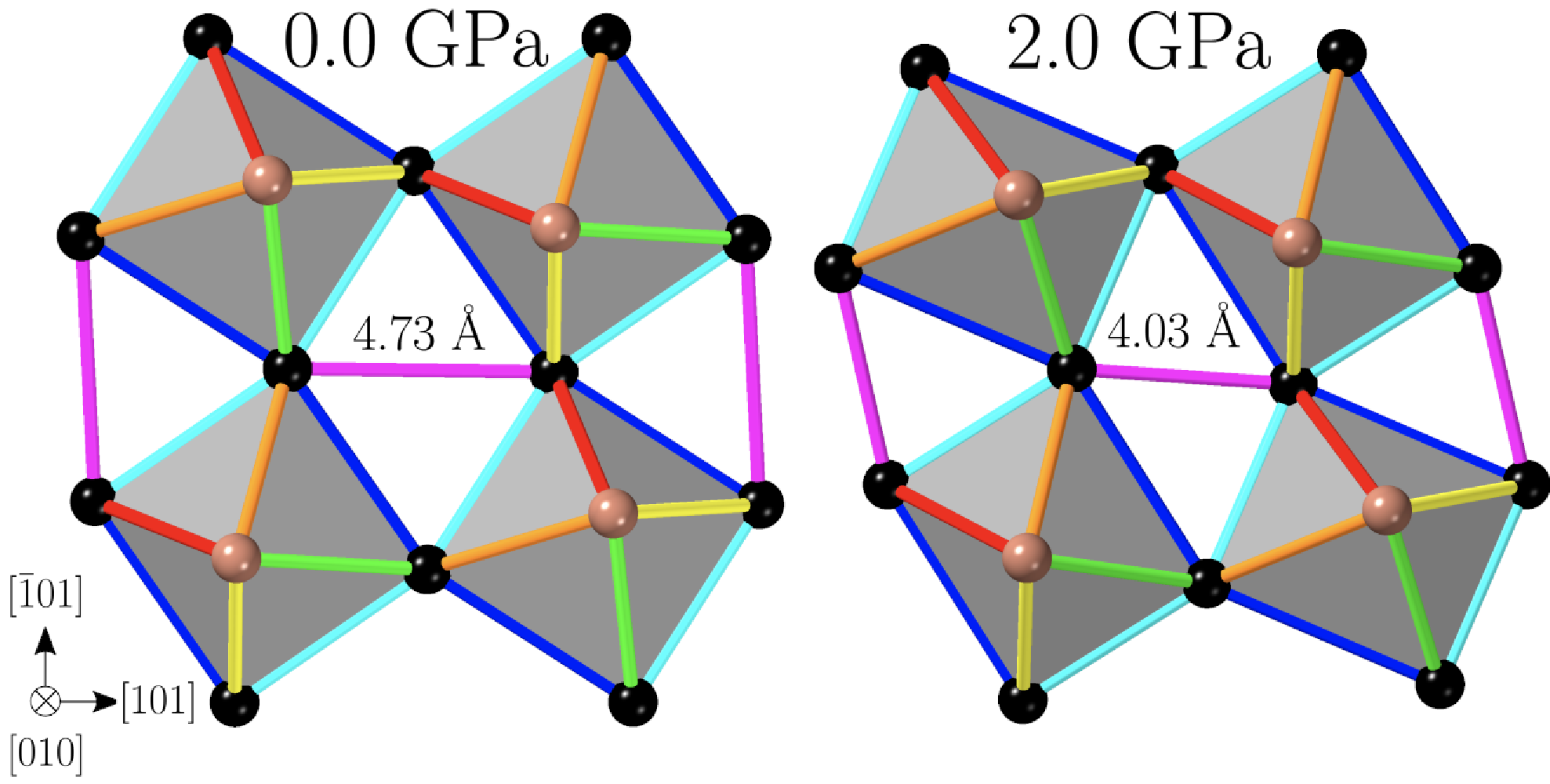}
    \caption*{}
    \label{fig:paths_NN_2x2_2_GPa}
\end{subfigure}%

\begin{subfigure}{0.9\textwidth}
    \centering
    \includegraphics[width=\linewidth]{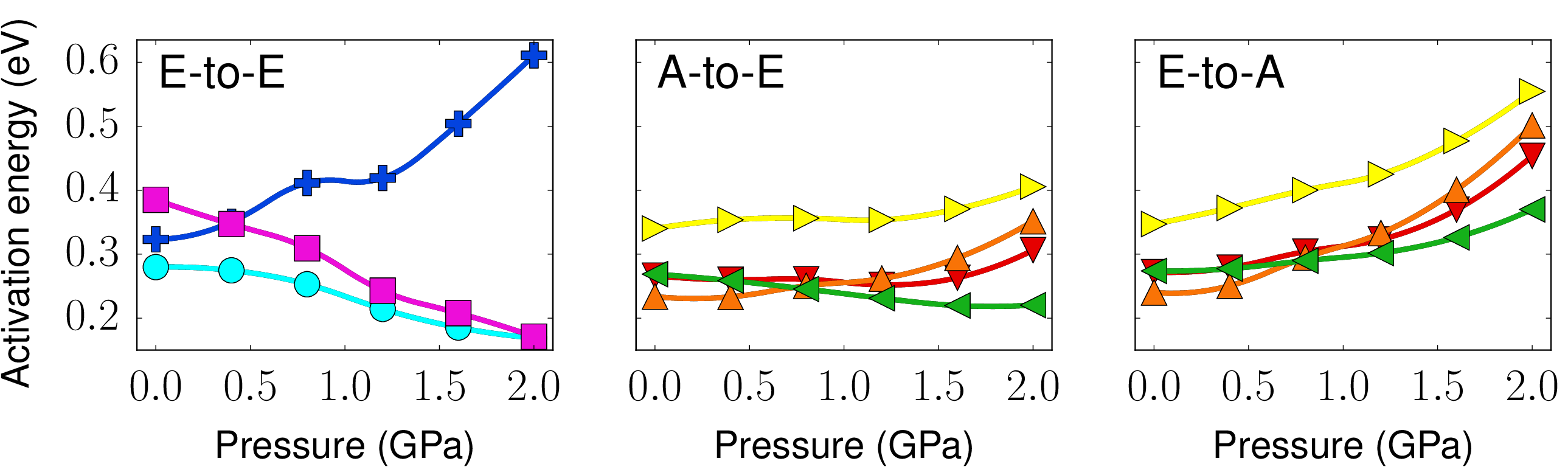}
    \caption*{}
    \label{fig:mig_bar_press}
\end{subfigure}%
\caption{Top row: Visual representation of the 2-NN pair of bromide sites, connected by a magenta stick, showing the large reduction in distance between them at 2.0 GPa (right) in comparison to 0.0 GPa (left). Bottom row; the energy barriers for vacancy migration under pressure for equatorial-to-equatorial (E-to-E), apical-to-equatorial (A-to-E) and equatorial-to-apical (E-to-A) paths, respectively.}
\label{fig:struc_NN_NNN_mig_bars}
\end{figure}

Any long range migration of vacancies requires multiple hops along symmetrically inequivalent elementary paths, so that the individual barriers are insufficient to draw meaningful quantitative conclusions; in particular, the connectivity of the lattice must be taken into account. In order to do so, we set up a kinetic scheme, described in detail in the computational methods, in which we account for the full connectivity of the lattice, and parametrise it using the activation barriers computed using DFT. We then use the kinetic scheme to compute the macrosopic mobility tensor $\boldsymbol{\mu}$, defined by the following equation:

\begin{equation}
    \vec{v}_{\mathrm{d}} = \boldsymbol{\mu} \vec{E}
\end{equation}
The eigenvectors of $\boldsymbol{\mu}$ are parallel to the lattice vectors, and the corresponding eigenvalues at 300 K (and 0.0 GPa) are 2.2, 4.8, 3.5 $\times$ 10$^{-6}$ cm$^2$V$^{-1}$s$^{-1}$ in the [100], [010] and [001] directions respectively. A decomposition of the elements into contributions from each of the symmetrically inequivalent paths, shown in the bottom right of Figure \ref{fig:disp_vecs_0_GPa}, indicates that multiple paths contribute to the mobility in each of the principal directions, leading to an essentially isotropic mobility; in contrast, taking any \emph{one} of the computed migration barriers, and calculating the mobility at 300K simply via the Arrhenius equation and Einstein relation, leads to values for the mobility which vary by two orders of magnitude, emphasising the importance of taking the full connectivity of the lattice into account. 
\\
We then start applying hydrostatic pressure, in the range of 0.0--2.0 GPa, to the $Pnma$ unit cell while constraining the space group, which is consistent with the experimentally and computationally observed stability range \cite{Zhang2017b,Nagaoka2017,Xiao2017c,Huang2019,Wang2019e}. Despite the pressure being isotropic, the structural response is significantly anisotropic, such that the length of \textbf{c} is reduced to about 90\% of its unstrained value at 2.0 GPa, in agreement with previous work \cite{Zhang2017b}. In contrast, the lengths of \textbf{a} and \textbf{b} remain largely unaltered, rising and falling, respectively, by approximately 1\%.
\\
In addition to the change in lattice vectors, the internal co-ordinates of the ions respond strongly to the applied pressure. This is mainly apparent in the equatorial (010) plane, where the octahedra exhibit enhanced tilting, leading to a structure described in previous publications as a ``squeezed wine-rack'' \cite{Zhang2017b}, as also observed in Refs. \cite{Xiao2017c,Nagaoka2017,Huang2019,Wang2019e}. As a result, there is a significant reduction in the distances between Br sites in neighbouring octahedra. One second-nearest-neighbour distance in particular, shown in Figure \ref{fig:struc_NN_NNN_mig_bars}, is significantly reduced under pressure, such that it approaches the range of values of 1-NN distances. Therefore, we also compute the activation energy for migration between that 2-NN pair of ions under pressure.

The evolution under pressure of the activation energies for migration between all seven pairs of sites under consideration is shown in Figure \ref{fig:struc_NN_NNN_mig_bars}. While most migration barriers increase with pressure, the barriers corresponding to two E-to-E transitions, including the 2-NN path discussed above, decrease such that they are significantly lower at 2.0 GPa, with values of 0.17 eV.

\begin{figure}[h]
\centering
\begin{subfigure}{0.36\textwidth}
\centering
    \includegraphics[width=\linewidth]{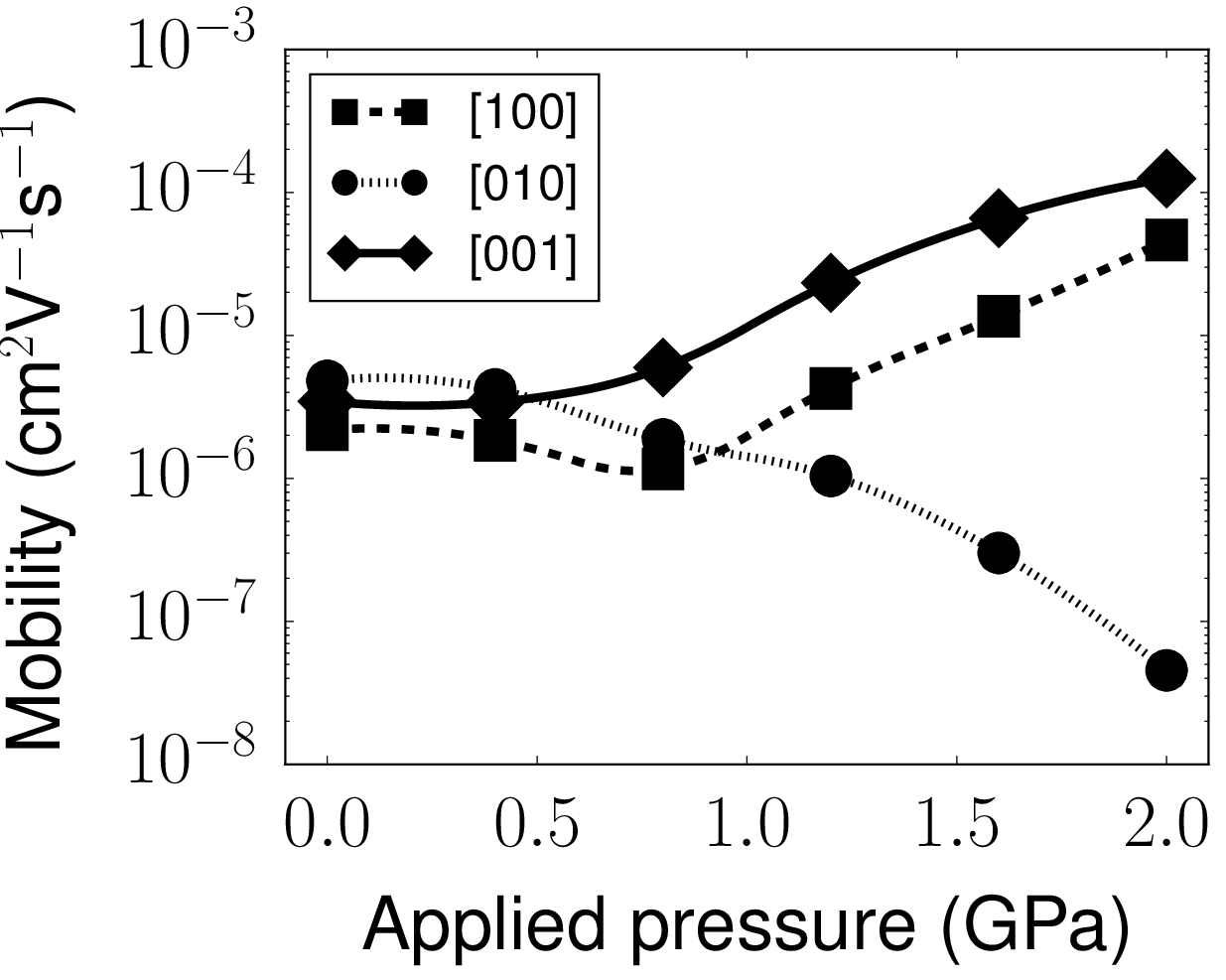}
    \caption*{}
    \label{fig:mob_tens}
\end{subfigure}%
\hfill
\begin{subfigure}{0.55\textwidth}
\centering
    \includegraphics[width=\linewidth]{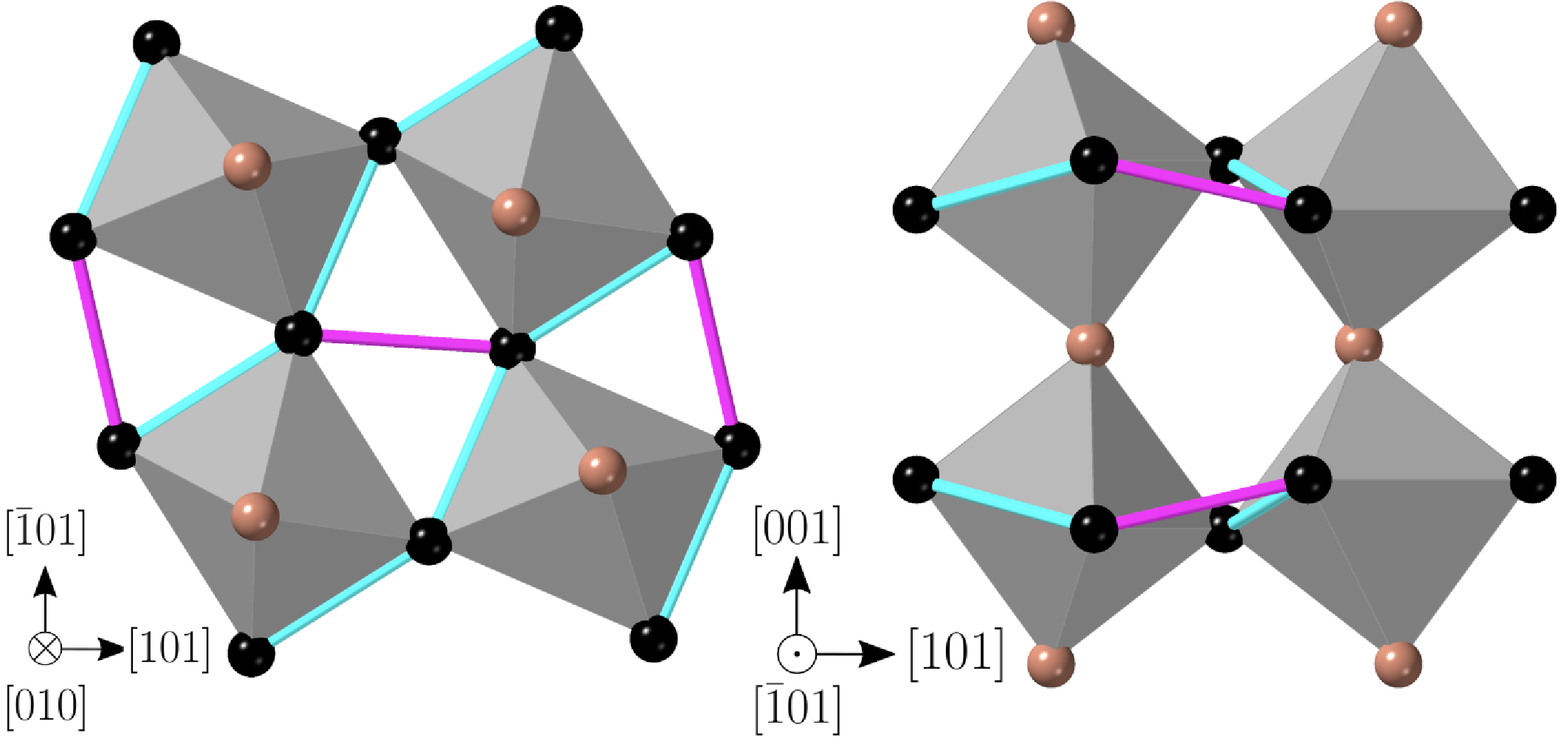}
    \caption*{}
    \label{fig:paths_2D_mob}
\end{subfigure}
\caption{Left: Dependence of the diagonal components of the mobility tensor on pressure, exhibiting an effective 3D-to-2D transition as pressure is increased. Centre and right: \ce{CsPbBr3} structure at 2.0 GPa from two different perspectives, showing the two pairs of lattice sites, the transitions between which contribute overwhelmingly to the mobility at that pressure.}
\label{fig:mob_tens_2_paths}
\end{figure}

The lowering of the energy barriers corresponding to the two paths has significant implications for the mobility at higher pressures, as the solution of the kinetic scheme shows. For all pressures, the eigenvectors of the mobility tensor remain along the crystalline [100], [010] and [001] directions; however, as shown in the left of Figure \ref{fig:mob_tens_2_paths}, the relative magnitudes of their eigenvalues change dramatically with increasing pressure such that at 2.0 GPa, the mobility in the equatorial $(010)$ plane is approximately 3 orders of magnitude higher than in the direction normal to it (parallel to $[010]$), leading to an effective 3D-to-2D transition in the mobility. A decomposition of the diagonal components of the mobility tensor in terms of the contributions of symmetrically inequivalent transitions shows that migration at 2.0 GPa is predominantly due to two transitions, namely the cyan and the magenta transitions shown in Figure \ref{fig:mob_tens_2_paths}, for which the migration barriers decrease significantly at elevated pressures.
\\

Due to the significant anisotropy of the mobility at elevated pressures, ion migration can either be strongly enhanced or strongly reduced depending on the orientation of the crystal with respect to the applied electric field. This effect could be exploited to control ion migration in layered devices, for example, where the lattice vectors and thermal coefficient of the LHP typically differ from that of the substrate. As a result, an intrinsic level of strain will be present at the interface between the perovskite film and the substrate, directly affecting the ionic migration in the vicinity of the interfaces even without using external pressure.

Finally, we note that the elastic constants of \ce{MAPbI3}, \ce{MAPbBr3} and \ce{CsPbI3} in the $Pnma$ phase show similar behaviour to \ce{CsPbBr3}, with c$_{22}$ being significantly higher than c$_{11}$ and c$_{33}$ \cite{Lee2018c}. As such, a similar anisotropic structural response to applied pressure might be expected, though this possibility, and its effects on ion migration, needs to be investigated further, particularly in light of the differences in bonding due to the organic molecule in the hybrid LHPs, and possible phase transitions under pressure.
\\
In conclusion, we have studied the effects of pressure on anion migration barriers and mobility, using a combination of DFT calculations and kinetic modelling. Our results demonstrate that a proper consideration of the point group symmetry and connectivity of the lattice structure of \ce{CsPbBr3} is necessary to model and understand anionic migration. At low pressures, anion migration proceeds via sequential transitions between multiple first-nearest neighbour (1-NN) pairs of sites, and is essentially isotropic. As pressure is increased, most of the barriers associated with transitions between 1-NN pairs increase, with only one decreasing significantly. Additionally, the activation energy for a transition between 2-NN halide sites is lowered significantly at higher pressures. The two effects combined lead to significantly enhanced mobilities in the equatorial plane, and a significantly reduced mobility normal to it, corresponding to an effective transition from 3D mobility in the absence of pressure, to effective 2D mobility at 2.0 GPa. This points to the potential of strain engineering perovskites to control the level of ionic migration in LHPs, in addition to their electronic properties. \\

\textbf{Methods: Density functional theory calculations} \\
Our DFT calculations were performed using the SCAN (strongly constrained and appropriately normed) meta-GGA exchange--correlation functional \cite{Sun2015b} and the PAW method for treating the effects of the atomic cores on the valence electrons \cite{Blochl1994}, as implemented in the Vienna \emph{ab initio} simulation package (VASP) code, version 5.4.4 \cite{Kresse1996}. We use the SCAN functional as it has been shown to outperform other exchange--correlation functionals in terms of the lattice parameters of perovskites \cite{Jia2019}, the local potential energy surface \cite{Bokdam2017}, the pressure-induced phase transitions \cite{Shahi2019} and lattice formation enthalpy \cite{Isaacs2018}.
\\
Optimisation of both the lattice parameters and the atomic coordinates was performed on the orthorhombic unit cell of CsPbBr$_3$, containing four formula units, under the constraint that the $Pnma$ space group of the structure was preserved, until both the energy change and change of eigenvalues between two self-consistent cycles were below 10$^{-8}$ eV and the residual forces were below 0.001 eV {\AA}$^{-1}$. The plane-wave cut-off energy was set to 600 eV. An $8 \times 6 \times 8$ $\Gamma$-centred $k$-point mesh was used, generated automatically according to the Monkhorst--Pack scheme \cite{Pack1976}.
\\
The calculations involving bromide vacancies were performed in a supercell of the orthorhombic phase, which was generated by using a [[2,0,2],[0,2,0],[-2,0,2]] scaling matrix on the $Pnma$ unit cell, and contains 320 atoms/64 formula units of the pristine lattice. Positively charged bromide vacancies were introduced in the cell, and, while keeping the lattice parameters fixed, internal coordinates were optimised using a plane-wave cutoff energy of 300 eV, evaluated at the $\Gamma$-point only, until both the energy change and change of eigenvalues between two self-consistent were below 10$^{-6}$ eV and the residual forces were below 0.01 eV {\AA}$^{-1}$.
\\
To determine the ion migration paths and the associated activation energies, climbing nudged elastic band (CNEB) calculations were performed as implemented in the VASP Transition State Tools (VTST) package \cite{Henkelman2000,Sheppard2008}. The migration path was evaluated using four structural images, with a spring constant of 5 eV/\AA$^2$ connecting the images. Internal coordinates were optimised using a plane-wave cutoff energy of 300 eV, evaluated at the $\Gamma$-point only, until both the energy change and change of eigenvalues between two self-consistent were below 10$^{-6}$ eV and the residual forces were below 0.05 eV {\AA}$^{-1}$. \\ \\

\textbf{Methods: Kinetic scheme and mobility tensor calculations} \\
To calculate the mobility tensor, we set up the kinetic scheme as a Continuous-Time Markov Chain, and consider a set of states of the system, each of which corresponds to a vacancy occupying a site in the unit cell \cite{Banyai1979,Allnatt2003}. The variables in the scheme are the probabilities of each state being occupied, the time dependence of which is given by solutions to a master equation:
\\
\begin{equation}
\frac{d\vec{P}}{dt} \ = \ \textbf{Q}\vec{P}.
\label{eq:dPdt}
\end{equation}
\\
In Equation \ref{eq:dPdt}, the elements of $\vec{P}$, $P_{j}$, are the probabilities of a vacancy occupying sites $j$ at time $t$, and the elements of the \emph{transition rate matrix} $\textbf{Q}$, $Q_{ij}$, are the rates (probabilities per unit time) for a vacancy to hop from site $j$ to site $i$. The transition rates, which are time independent in our scheme, are calculated using the Arrhenius expression:
\\
\begin{equation}
\label{eq:exp_rate}
    Q_{ij} = A \exp{\left(-\beta (E_{\mathrm{a}})_{ij}\right)},
\end{equation}
\\
in which $\beta = \left(k_{\mathrm{B}}T\right)^{-1}$, $(E_{\mathrm{a}})_{ij}$ is the activation energy for a vacancy to hop from site $j$ to site $i$, and $A$ is the attempt frequency for the process to take place, which we assume to be the same for all pairs $i$ and $j$. In our calculations, we assume a temperature $T$ of 300 K and an attempt frequency $\nu$ of 2$\times$10$^{12}$ s$^{-1}$, which we approximate based on the Raman frequency of 70 cm$^{-1}$ corresponding to the inorganic \ce{PbBr6} cage \cite{Zhang2017b,Xue2019}, and which we assume is constant with pressure \cite{Zhang2017b}, in accordance with earlier work on LHPs \cite{Muscarella2020}. $(E_{\mathrm{a}})_{ij}$ is taken as the sum of the result of the DFT CNEB calculation described above, which we denote $(E_{\mathrm{a}}^{0})_{ij}$, and a field dependent part, which we calculate simply as the difference in potential energy between site $j$ and the location of the moving anion at the (energetic) saddle point of its path between sites $i$ and $j$ (which corresponds to the hopping of the vacancy between sites $j$ and $i$):
\\
\begin{equation*}
    (E_{\mathrm{a}})_{ij} = (E_{\mathrm{a}}^{0})_{ij} + q \left(\vec{r}_{\mathrm{saddle}}-\vec{r}_{j}\right) \cdot \vec{E}
\end{equation*}
where we use the formal charge $q$ of -1.0 for the moving anion \cite{DeSouza2020}. We note that this approach is a generalisation of the approach adopted originally by Mott and Gurney \cite{Mott1950,Genreith-Schriever2016,Salles2020}. Further, we stress that the internal electric field $\vec{E}_{\mathrm{int}}$ will differ from the applied field $\vec{E}$ and will depend strongly on the experimental conditions, as such we do not account for it in our model. 

In order to compute the mobility, we ``apply'' a finite electric field, and then solve Equation \ref{eq:dPdt} in the steady state:
\\
\begin{equation}
\frac{d\vec{P}}{dt} \ = \vec{0} = \ \textbf{Q}\vec{P}.
\label{eq:dPdt_ss}
\end{equation}
\\
We can do so by finding the eigenvectors of the rate matrix, with the one with eigenvalue 0 corresponding to steady state. We note that at equilibrium (i.e. zero field), they would have values given by Boltzmann factors of the vacancy formation energy, but when a field is applied that is no longer the case.
\\
Once we have the steady state probabilities $P_{j}$, we can compute the flux from state $j$ into $i$---that is, the number of hopping events of a vacancy between those sites, per unit time---is $Q_{ij}P_{j}$, and finally the drift velocity $\vec{v}_{\mathrm{d}}$ as the sum of the fluxes multiplied by the displacement vectors associated with each hop, $\vec{r}_{i}-\vec{r}_{j}$:
\\
\begin{equation}
     \vec{v}_{\mathrm{d}} = \sum_{j} \sum_{i\neq j} Q_{ij} P_{j} (\vec{r}_{i}-\vec{r}_{j})
\label{eq:drift_sum}
\end{equation}
\\
To obtain the components of the full mobility tensor $\mu$, we apply a field along the crystalline axes [100], [010] and [001] in turn, and then compute $\vec{v}_{\mathrm{d}}$ by summing over all hops. In general, one can diagonalise the tensor matrix, in order to obtain the mobility eigenvalues and the corresponding directions. In our study, we find that the eigenvectors of the mobility tensor are parallel to the lattice vectors.
\\

\begin{acknowledgement}

We acknowledge funding from the European Union’s Horizon 2020 MSCA Innovative Training Network under grant agreement number 764787, and the Energy Oriented Centre of Excellence (EoCoE-II), grant agreement number 676629. This research made use of the Balena High Performance Computing (HPC) Service at the University of Bath; and the Isambard UK National Tier-2 HPC Service (http://gw4.ac.uk/isambard/) operated by GW4 and the UK Meteorological Office, and funded by EPSRC (EP/P020224/1). We thank Drs. I. R. Thompson and W. R. Saunders for discussions pertaining to kinetic modelling, and Dr. T. Ducho\v{n} for commenting on the manuscript. 
\end{acknowledgement}

\bibliography{library}

\end{document}